\documentclass[aps,prl,twocolumn,letterpaper,superscriptaddress,showpacs]{revtex4}
\usepackage{graphicx}
\usepackage{CJK}

\begin{document}

\begin{CJK*}{UTF8}{bsmi}
\title{Weak phase stiffness and nature of the quantum critical point in underdoped cuprates}
\author{Yucel Yildirim}
\altaffiliation{current address: Physics Department, Faculty of Arts and Sciences, Do\u{g}u\c{s} University, Ac{\i}badem-Kad{\i}k\"{o}y, 34722 Istanbul, Turkey}
\affiliation{CMPMSD, Brookhaven National Laboratory, Upton, NY 11973-5000,U.S.A.}

\author{Wei Ku (\CJKfamily{bsmi}顧威)
}
\altaffiliation{corresponding email: weiku@bnl.gov}
\affiliation{CMPMSD, Brookhaven National Laboratory, Upton, NY 11973-5000,U.S.A.}
\affiliation{Physics Department, State University of New York, Stony Brook, New York 11790, USA}
\date{\today}

\begin{abstract}
We demonstrate that the zero-temperature superconducting phase diagram of underdoped cuprates can be quantitatively understood in the strong binding limit, using only the experimental spectral function of the ``normal'' pseudo-gap phase without any free parameter.
In the prototypical (La$_{1-x}$Sr$_x$)$_2$CuO$_4$, a kinetics-driven $d$-wave superconductivity is obtained above the critical doping $\delta_c\sim 5.2\%$, below which complete loss of superfluidity results from local quantum fluctuation involving local $p$-wave pairs.
Near the critical doping, a enormous mass enhancement of the local pairs is found responsible for the observed rapid decrease of phase stiffness.
Finally, a striking mass divergence is predicted at $\delta_c$ that dictates the occurrence of the observed quantum critical point and the abrupt suppression of the Nernst effects in the nearby region.
\end{abstract}


\maketitle
\end{CJK*}

Considering the enormous amount of research activities devoted to the problem of high-$T_c$ superconductivity, it is hardly an exaggeration to regard it as one of today's most important unsolved problems in physics.
Specifically in the underdoped region of cuprates, it is now commonly accepted that the low carrier density in the system necessarily leads to strong phase fluctuation of the superconducting order parameter\cite{DoniachInui,EK} due to its conjugate nature to the number fluctuation.
Consequently, the transition temperature $T_c$ is suppressed significantly below the pairing energy scale that controls all essential aspects of the standard theory of superconductivity\cite{BCS}.
The crucial role of phase fluctuation\cite{EK,orderpar2,ChargedLatticeBoson} has recently gained strong support from various experiments\cite{Samuele,Broun,Hardy,debate} in both the low-temperature superconducting state and the `normal state' above the transition temperature $T_c$, and is likely tied closely to many of the exotic properties\cite{EK,UD4,UD5,UD6,UD8,UD10} in this region.

Nonetheless, besides this general understanding, several key issues remain puzzling in the underdoped region.
In spite of an uneventful evolution of the one-particle spectral function\cite{UD8}, the superfluid density reduces \textit{dramatically} near the observed quantum critical point (QCP)\cite{QCP} (at the critical doping $\delta_c\sim 5.2\%$ for doped La$_2$CuO$_4$), below which superconductivity ceases to exist even at zero temperature.
The current consideration of phase fluctuation~\cite{EK} would only indicate a softer phase at lower carrier density, but offers no explanation for the complete suppression of superconductivity at zero temperature at $\delta < \delta_c$.
Particularly in La$_2$CuO$_4$, $\delta_c$ is quite far away from the antiferromagnetic (AF) phase boundary, rendering the common consideration of competing order unsatisfactory.
This vanishing of superconductivity below $\delta_c$, the nature of the QCP, the dramatic reduction of superfluid density nearby, and the controlling factor of the value of $\delta_c$, all remain challenging to our basic understanding.

Perhaps the most puzzling observation is the sudden suppression of the observed Nernst effect at $T > T_c$ around the same critical doping $\delta_c$\cite{OngNernstEffect&Diamagnetism}.
This indicates that not only the long-range phase coherence, but also the shorter-range phase coherence is lost near the QCP, a phenomenon unexplainable via simple fluctuation scenario, for example due to low dimensionality.

In this letter, we demonstrate that these puzzles can be quantitatively understood in the strong binding limit of local pairs of doped holes.
We obtain the zero-temperature underdoped phase diagram with no need for any free parameter, other than the experimental one-particle spectral function of the pseudo-gap ``normal'' state.
A kinetics-driven $d$-wave condensate is found at $\delta > \delta_c$, with a largely enhanced bosonic mass, $m^* > 40m_e$.
In great contrast, ground states consisting of fluctuating $p$-wave pairs are found at $\delta < \delta_c$, incapable of sustaining a condensate.
At $\delta=\delta_c$, a mass divergence results from the degeneracy of local $d$- and $p$-wave symmetry, dictating the presence of the QCP.
Correspondingly, near the QCP $\delta\ge\delta_c$, the diverging mass explains the puzzling dramatic reduction of phase stiffness in both long range and shorter range.
Our study provides a novel yet simple paradigm to the behavior of local pairs in underdoped cuprates, and is expected to inspire new set of experimental confirmation, as well as re-interpretation of existing experimental observations.

Conceptually, a phase-fluctuation dominant superconductivity hosts relatively negligible amplitude fluctuation of the order parameter at low energy/temperature.
This implies that the effective low-energy Hamiltonian for the charge and pairing channels must have integrated out all pair-breaking processes to conserve the amplitude of the order parameter, for example, as in the $x$-$y$ model\cite{xyModel}.
The higher-energy pairing scale should then manifest itself only through a strong ``pair-preserving'' constraint of the low-energy Hamiltonian.
This is in perfect analogy to the replacement of repulsion $U$ of the Hubbard model by a ``no double occupancy'' constraint in its lower energy counterparts, say the $t$-$J$ model.
Consequently, a new paradigm for the low-energy physics emerges at $T\leq T_c$, which differs completely from the emphasis of amplitude fluctuation in the standard theories.
In this new physical regime, the detail of the pairing mechanisms (AF correlation\cite{ChargedLatticeBoson,AndersonGlue,SlaveBoson,AFcorr1,AFcorr2}, spin-fluctuation\cite{SpinFluct}, or formation of bi-polaron\cite{biPolaron}) are no longer essential.
Instead, the physics is now dominated by the effective kinetic energy that controls the phase coherence.
Since only one energy scale is essential in this regime, the low-energy physics should be universal and simple.

Below, we proceed to 1) obtain the effective kinetics of the doped holes from the experimental one-particle spectral function in the ``normal state'' pseudogap phase, 2) derive the effective motion of tightly-bound pairs of holes under the pair-preserving constraint, and 3) solve the resulting bosonic problem to address the physical issues quantitatively without any free parameter.

\begin{figure}
\includegraphics[width=8.5cm]{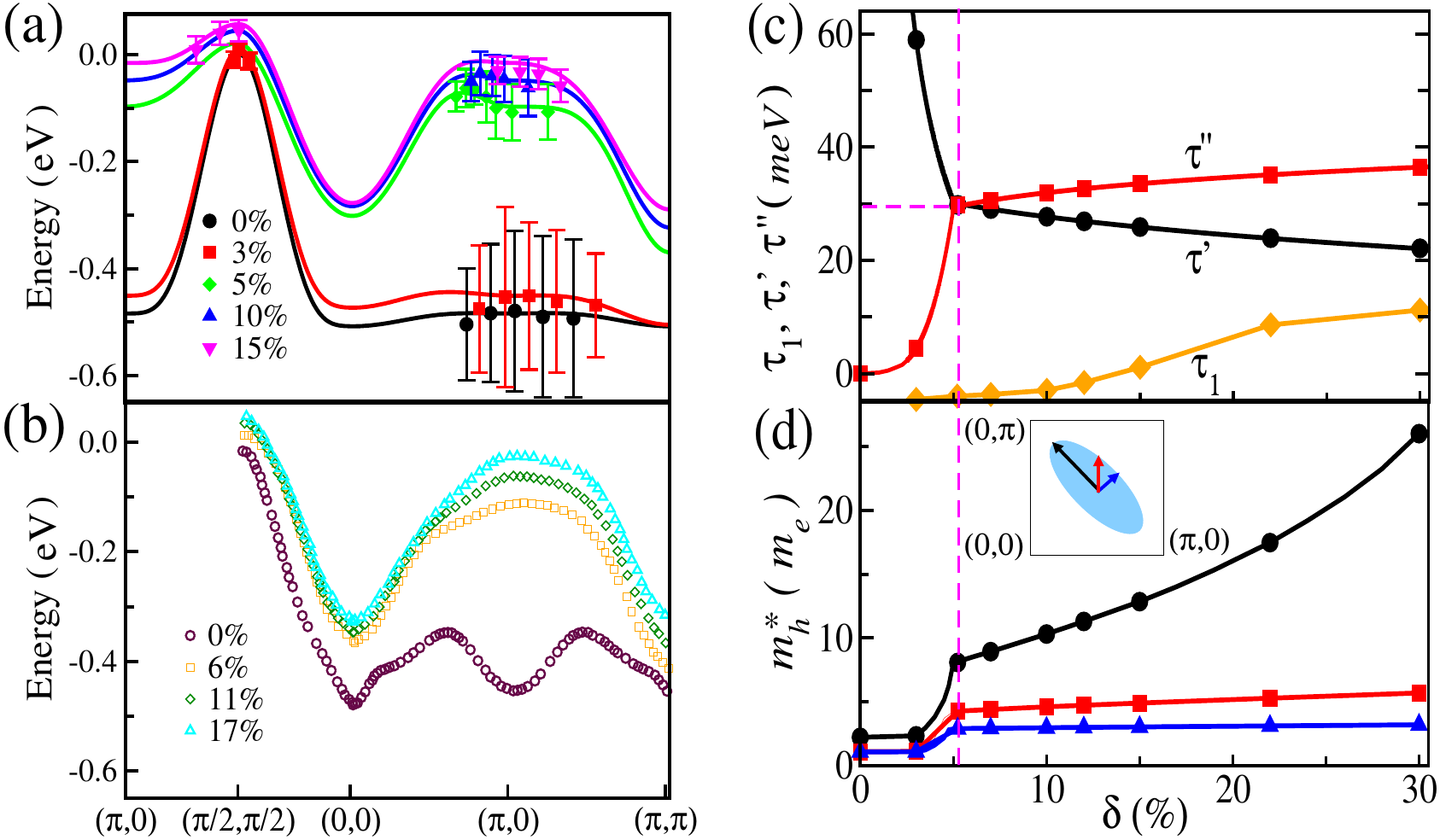}
\caption{\label{fig:fig1} 
(Color online) Doping dependent band dispersion obtained from experiment\cite{ExpArpes1-2,ExpArpes3-4} [dots in (a)], Eq.(\ref{eq:eqn2}) [lines in (a)], and the $t$-$J$ model\cite{Weiguo} (b), with chemical potential at zero.
(c) Corresponding $\tau$, $\tau^\prime$ and $\tau^{\prime\prime}$ in the \textit{hole} picture.
(d) Doping dependence of the effective mass of holes $m^*_h$ in the major directions indicated by the arrows in the inset.}
\end{figure}

1) \textit{Effective kinetics}
The dots in Fig.~\ref{fig:fig1}(a) gives the dispersion of the main features in the experimental spectral functions of the ''normal state'' of (La$_{1-x}$Sr$_x$)$_2$CuO$_4$ in the pseudogap phase, obtained by angular-resolved photoemission spectroscopy (ARPES)\cite{ExpArpes1-2,ExpArpes3-4}.
One notices immediately that the dispersion is strongly doping ($\delta$) dependent, especially near ($\pi$,0).
Judging from the close resemblance to the published $t$-$J$ model solutions\cite{Weiguo} in Fig.~\ref{fig:fig1}(b), this strong band renormalization likely originates from the competition between the bare kinetic energy and the AF interaction.~\cite{pseudogap}
The effective kinetics of carriers can then be captured by the irreducible kinetic kernel $\tau \equiv G_L^{-1} - G^{-1}$ (in matrix notation and in the \textit{hole} picture) through the measured one-particle propagator $G$ and a reference non-propagating Green's function $G_L$, defined with a single pole at the central energy of the band.
The real part of the off-site elements of $\tau$ thus controls the propagation of the carriers, just like the effective hopping matrix elements.
The imaginary part of $\tau$ gives the decay of carriers and becomes large at $\omega > 0.3$ eV where the spectral function is broad and quasiparticle description no longer applies.
Since only the average motion at long time scale is of significance in this study, we will drop the imaginary part and represent the average kinetics via
\begin{equation}
\label{eq:eqn2}
H = \sum_{ii^\prime} \tau_{ii^\prime}c_i^\dagger c_{i^\prime} + h.c.
\end{equation}
for simplicity~\cite{supp}.
In this case, $\tau_{ii^\prime}$ is equivalent to those from a tight-binding fit of the experimental dispersion.

Note that this Hamiltonian is only meant to capture the average effective kinetics of the fully renormalized one-particle propagator.
It does not contain information of the pairing interaction that connects to the high-energy sector.
The use of Hamiltonian representation here is merely for better clarity of the underlying physics~\cite{supp}.
Furthermore, $\tau$ is to be distinguished from the ``bare'' hopping parameter $t$ commonly used in the Hubbard or $t$-$J$ model, as $\tau$ have fully absorbed the effects of interactions and constraints.
Finally, the actual carriers do \textit{not} need to be quasi-particles, and their ''diffusive'' nature near ($\pi$,0) can be included by keeping the full $\tau$ in the study~\cite{supp,T_pair}, and all our physical conclusions below would remain.

The resulting doping dependent first, second and third neighbor kernels, $\tau_1$, $\tau^\prime$ and $\tau^{\prime\prime}$, are shown in Figure~\ref{fig:fig1}(c), and correspond to dispersion curves [lines in Fig.~\ref{fig:fig1}(a)] comparable to the experimental ones.
Interestingly, as $\delta$ decreases, $\tau^{\prime\prime}$ is found to increase steadily approaching the value of $\tau^\prime$, and then exceeds $\tau^\prime$ {\it right at $\delta_c$}!
This is apparently not a coincidence, and reveals an important clue to the nature of the QCP to be discussed below.
Due to the strong AF correlation, the fully dressed $\tau_1$ is negligibly small at the underdoped regime and will be dropped from our further analysis.
As a reference, Fig.~\ref{fig:fig1}(d) also shows a weakly doping dependent effective mass of the doped holes, $m^*_h$, for $\delta>5.2\%$ in three major directions, consistent with the current lore\cite{EffectiveMass}.

2) \textit{Motion of Tightly-Bound Pairs}
Since it is unlikely that doped holes can doubly occupy the same site in a weakly doped AF Mott insulator, it is reasonable to assume that under a strong binding, pairs mostly consist of nearest neighboring holes.
It is thus convenient to employ a bosonic representation of pairs, $b_{ij}^\dagger=c_{i \uparrow}^\dagger c_{j \downarrow}^\dagger$, located at neighboring site $i$ and $j$ with opposite spin.
Such a real-space hole pair can result from numerous high-energy mechanisms\cite{ChargedLatticeBoson,AFcorr1,AFcorr2,biPolaron}, and is to be distinguished from the real-space \textit{singlet} pair of \textit{electrons} in RVB-like constructions\cite{RVB}.

\begin{figure}
\includegraphics[width=8.5cm]{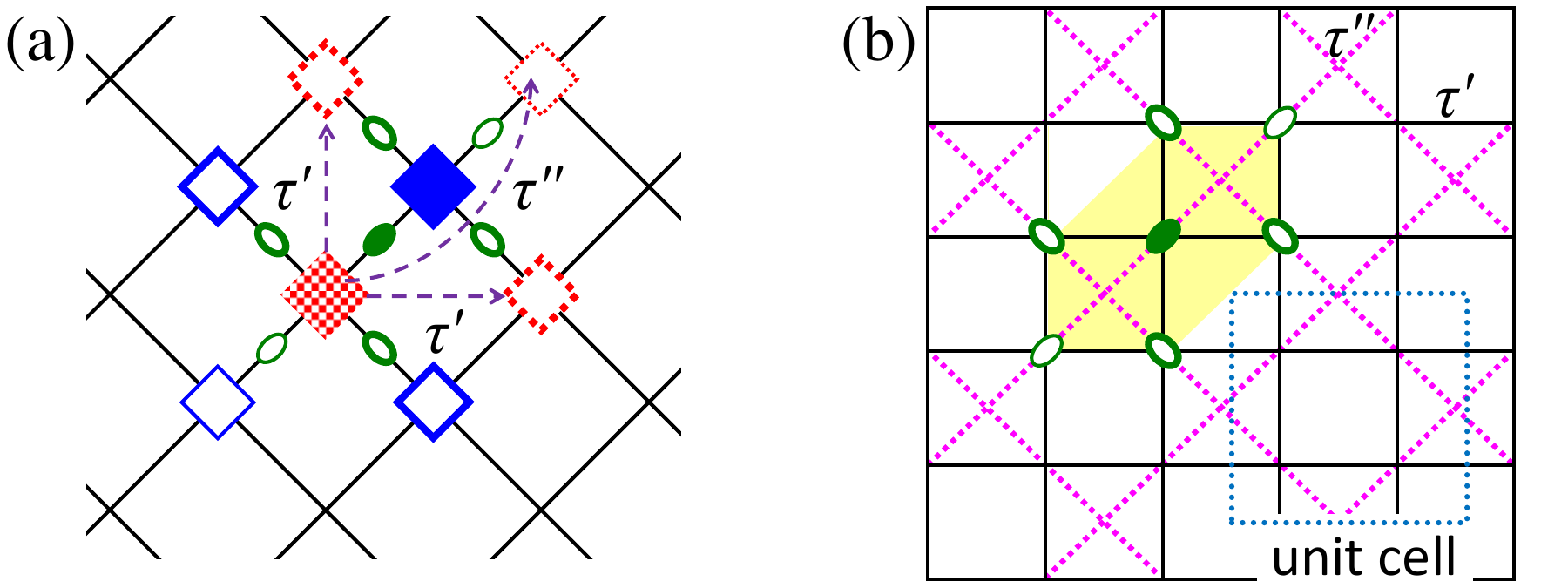}
\caption{\label{fig:fig2} 
(Color online)  Illustration of kinetic processes of a pair of holes (filled diamond) to its six allowed destinations (open diamonds) under the ``pair-preserving'' constraint (a), through $\tau^\prime$ (solid lines) and $\tau^{\prime\prime}$ (dashed lines).
The same is equivalently represented by ellipsoids denoting a pair and its allowed six neighbors (b).
The yellow area denotes the `extended hardcore constraint' that excludes other pairs.
}
\end{figure}

Now, consider the motion of a single pair of holes (blue and red filled diamonds) located in the fermion lattice in Fig.~\ref{fig:fig2}(a).
Under the pair-preserving constraint, only three potential destinations (empty diamonds) for each hole are allowed, two via second neighbor hopping, $\tau^\prime$, one via third neighbor hopping, $\tau^{\prime\prime}$.
Converting to the lattice of bond-centered pairs in Fig.~\ref{fig:fig2}(b), one finds a checkerboard lattice consisting of two nonequivalent sites, each connecting to four first neighboring sites via $\tau^\prime$, but to only \textit{two} second neighboring sites via $\tau^{\prime\prime}$.
This pivoting motion of the paired holes can then be represented by
\begin{equation}
\label{eq:eqn4}
H^b = \sum_{ii^\prime j} \tau_{ii^\prime} b_{ij}^\dagger b_{i^\prime j} + h.c.
\end{equation}
The same motion was previously derived via a rigorous separation of many-body Hilbert subspace of paired holes~\cite{QPGap}.
Optionally, one can also include both the real and imaginary part of $\tau$ via the equation of motion, or the ladder diagrams~\cite{supp,T_pair}.
Although, inclusion of the imaginary part of $\tau$ introduces broadening of the bosonic propagator at higher energy, but has little effect on the condensation taking place at low energy.

Note that the hole pairs $b$'s are under a strong `extended hardcore constraint': $b_{ij}^\dagger b_{i'j'}^\dagger=0$ if $i=i'$ or $j=j'$.
This is inherited from the Pauli exclusion principle of the original fermion operators and that double occupancy of electrons are not allowed in the low-energy sector.
Indicated by the yellow area in Fig.~\ref{fig:fig2}(b), this constraint forbids occupation by another pair at any of the six potential hopping destinations of a pair.
It can be considered as an infinite short-range repulsion that determines the bare scattering length between pairs, and is responsible for stabilizing the bosonic system against phase separation\cite{NoPhaseSep}.

\begin{figure}
\includegraphics[width=8.5cm]{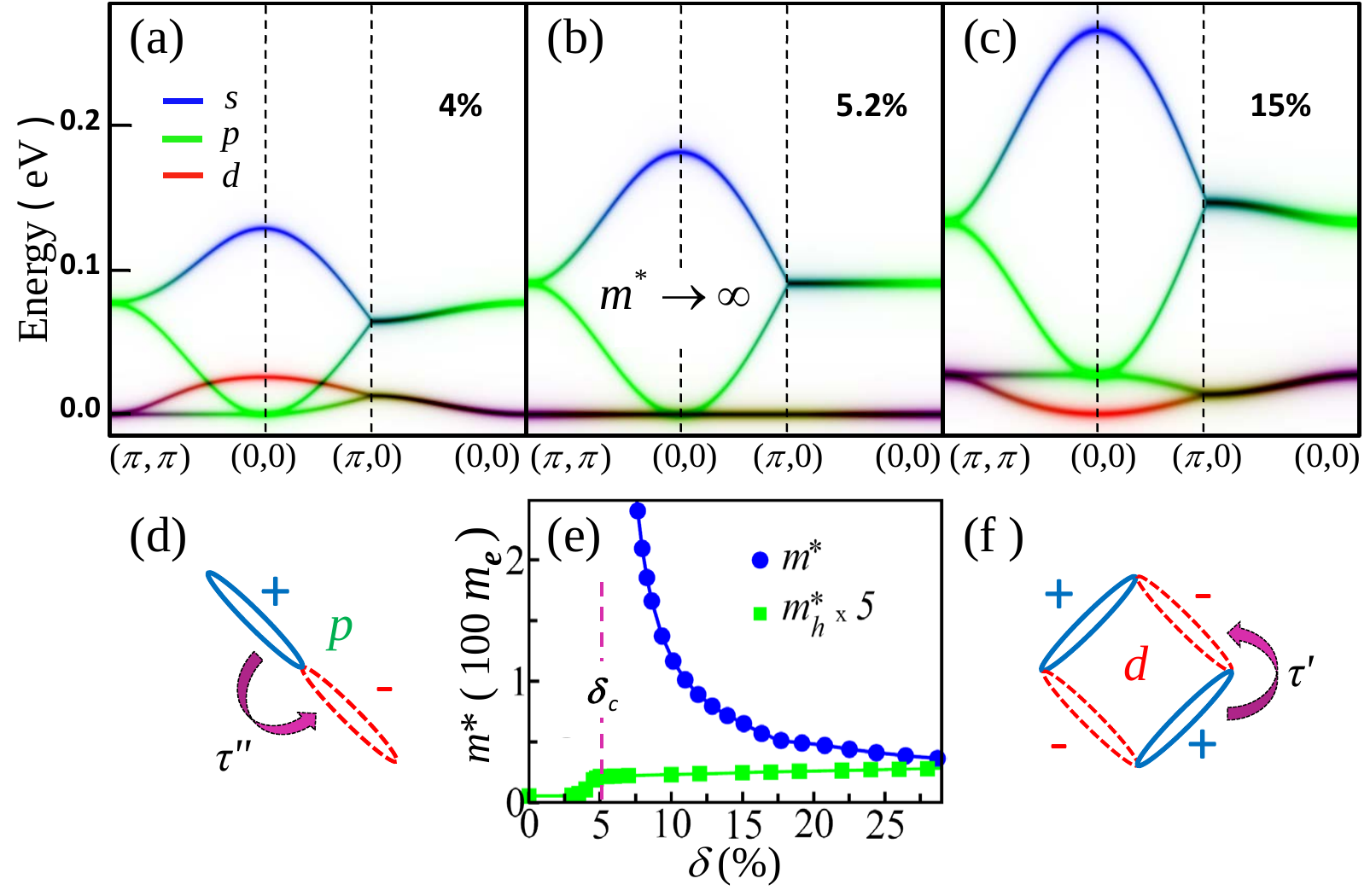}
\caption{\label{fig:fig3} 
(Color online) The band dispersion of the hole pairs without the extended hardcore constraint, at $\delta<\delta_c$(a), $\delta=\delta_c$(b), and $\delta>\delta_c$(c).
(d) and (f) illustrates the dominant kinetic process and the Wannier function corresponding to the lowest band in (a) and (c) respectively.
(e) Strongly enhanced effective mass of the pairs $m^*$ and the mass of the holes, $m^*_h$.}
\end{figure}

3) \textit{Results}
We diagonalize Eq.(\ref{eq:eqn4}) first without the extended hardcore constraint, using a unit cell containing four sites shown in Fig.~\ref{fig:fig2}(b).
This choice explicitly allows one $s$-, two $p$-, and one $d$-wave superposition within the unit cell, and equates the doping level per unit cell in this lattice and that in the standard fermion lattice.
Fig.~\ref{fig:fig3}(c) illustrates the resulting band structure in the superconducting phase at doping $\delta=15\% > \delta_c$.
It shows that at low enough temperature a Bose-Einstein condensate (BEC) would take place at a single minimum at momentum $q=0$, with a pure $d$-wave symmetry (red color).
As in standard dilute bosonic systems, one thus expects a $d$-wave superfluid with finite stiffness, once a scattering length (derived primarily from the extended hardcore constraint) is switched on.

The local $d$-wave structure of the pair is better illustrated in real space via the corresponding Wannier function in Fig.~\ref{fig:fig3}(f), computed from the Fourier transform of the Bloch functions of the lowest band.
The low-energy pairs has clear $d$-wave symmetry with nodes along the $(\pi,\pi)$ directions of the standard Fermion lattice, in perfect agreement with the experimental observations.\cite{UD8,arpes2,arpes3}
(Notice in Fig.~\ref{fig:fig2}(a) that our fermionic lattice is rotated by $45^\circ$ from the usual convention.)

We stress that our resulting local $d$-wave symmetry is completely driven by the {\it fully screened kinetic energy}.\cite{t1J_Dagotto,BareKinetic}
It originates from the dominance of positive $\tau^\prime$ of the local pair, which prefers energetically opposite sign of the wave function across first neighbors, thus favoring a $d$-wave symmetry [see Fig.~\ref{fig:fig3}(f)].
In comparison, the positive $\tau^{\prime\prime}$ favors opposite sign across the second neighbors, thus $p$-wave symmetry [see Fig.~\ref{fig:fig3}(d)].
Therefore, $\tau^\prime$ and $\tau^{\prime\prime}$ compete by lowering the band energy of $d$- and $p$-bands, respectively.

This explains the long-standing puzzle of lack of superconductivity at lower doping ($\delta < \delta_c$).
Since in this region $\tau^{\prime\prime} > \tau^\prime$ [c.f. Fig.~\ref{fig:fig1}(c)], Fig.~\ref{fig:fig3}(a) shows that local $p$-wave pair has lower energy than $d$-wave pairs.
Furthermore, in the checkerboard lattice in Fig.~\ref{fig:fig2}(b), the parity of $p$-states dictates a line of degeneracy (green flat band in Fig.~\ref{fig:fig3}) from (0,0) to ($\pi$,$\pi$).
The pairs can therefore populate any arbitrary state along this line without ever forming a BEC.
The system is thus composed of incoherent $p$-wave pairs, an effect of quantum phase fluctuation beyond the original consideration of thermal phase fluctuation~\cite{EK}.

The competition between $d$-wave and $p$-wave also offers a natural explanation of the dramatic phase softness and the low superfluid density of the underdoped cuprates.
Indeed, even near the optimal doping ($\delta\approx15\%$), the comparable value of $\tau^{\prime\prime}$ and $\tau^\prime$ leads to a large effective mass of the pair $m^*=(\hbar^2/l^2) d^2\epsilon_k/dk^2 \approx 12m^*_h \approx 59m_e$ ($l$ being the lattice constant).
This gives a rather long penetration depth $\lambda=\sqrt{{{m^*c^2}\over {4\pi e^2 n_s}}} \approx 7000 \AA$ (taking $n_s\sim \delta$ per unit cell), in reasonable agreement with the experimental value \cite{LambdaExp}.
Furthermore, as $\delta$ decrease toward $\delta_c$, $\tau^{\prime\prime}$ grows to the value of $\tau^\prime$, reducing the separation of the $d$-band and the $p$-band, and in turn flattening the $d$-band.
The effective mass of the $d$ band thus increases significantly (Fig.~\ref{fig:fig3}(e)), consequently giving rise to the observed very small phase stiffness.

This analysis reveals the simple yet exotic nature of the observed QCP at the end of the underdoped superconductivity region $\delta_c=5.2\%$: It is dictated by the diverging effective mass of the local pairs [Fig.~\ref{fig:fig3}(e)].
At this point $\tau^\prime=\tau^{\prime\prime}$ and the $d$-wave and $p$-wave pairs become locally degenerate and the $d$-band is thus completely flat, as shown in Fig.~\ref{fig:fig3}(b).
Since the effective mass now diverges, the pairs can no longer propagate and align the phase to develop a condensate.
In essence, it is the perfect quantum interference between $\tau^\prime$ and $\tau^{\prime\prime}$ that renders the local pairs immobile, and in turn disables the phase coherence of superconductivity.

This result also explains nicely the puzzling dramatic suppression of diamagnetic response~\cite{dia1} and Nernst signal~\cite{OngNernstEffect&Diamagnetism} near $\delta_c$.
Indeed, within phase fluctuation scenario, a divergent mass might be the only way to completely suppress the shorter-range coherence responsible for a strong diamagnetic response.

\begin{figure}
\includegraphics[width=8.5cm]{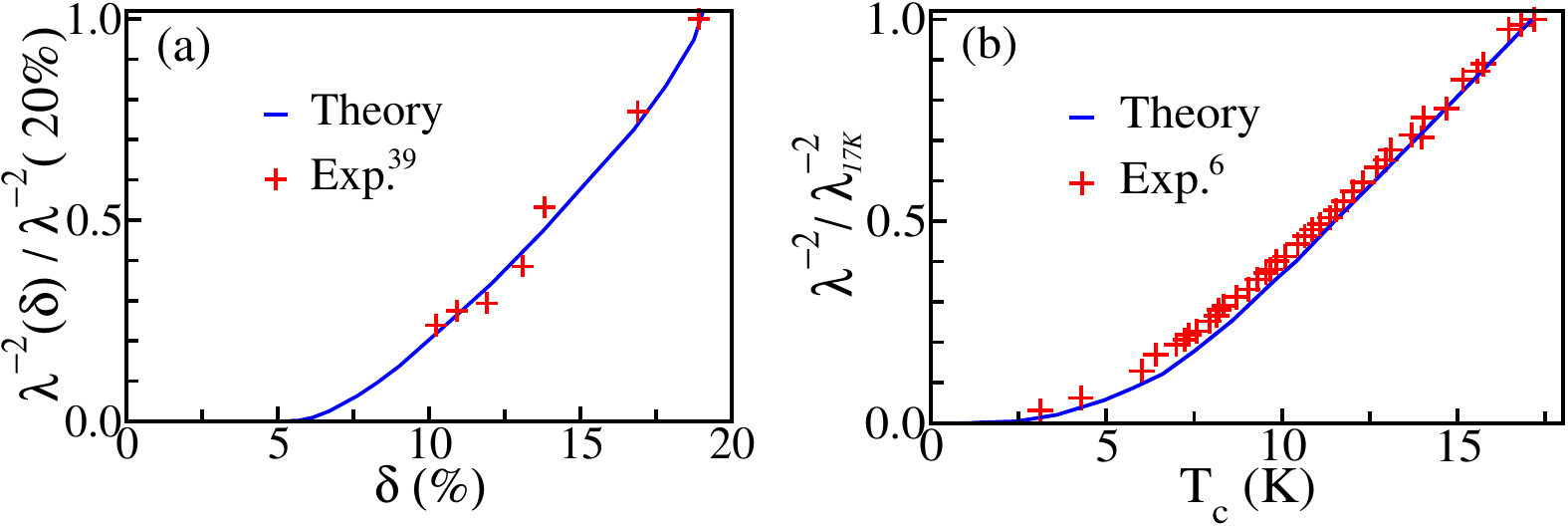}
\caption{\label{fig:fig4} 
(Color online)  Experimental supports of predicted mass divergence via (a) non-linear doping dependence of inverse penetration depth and (b) non-linear correlation between inverse penetration depth and transition temperature $T_c$.}
\end{figure}

Our predicted mass divergence near QCP is actually strongly supported by experimental measurements of penetration depth $\lambda$ of the underdoped YBa$_2$Cu$_3$O$_y$ samples.
Figure~\ref{fig:fig4}(a) shows that over the entire underdoped region, the measured $\lambda^{-2}$~\cite{Sonier} deviates significantly from the simple $\lambda^{-2}\propto\delta$ relationship to be expected with a constant effective mass.
On the other hand, our theory with large doping-dependent effective mass reproduces very nicely the experimental observation.
A even more direct evidence is provided by the recent measurement on the extremely underdoped YBa$_2$Cu$_3$O$_y$ samples near the QCP~\cite{Broun}.
The observed relationship between low-temperature $\lambda^{-2}$ and $T_c$ in Fig.~\ref{fig:fig4}(b) shows a strong non-linear dependence.
In fact, the same behavior has also been observed via mutual inductance\cite{DifferentExp}.
The zero slope at $\lambda\rightarrow 0$ can be interpreted as an indirect evidence of the mass divergence, and our theory reproduces very nicely the experimental observation~\cite{Tc_determination}.


Our analysis has wide scope of implications in the electronic structure of the underdoped cuprates that deserve further investigations.
As $\delta$ decreases toward $\delta_c$, the diverging mass makes perfect sense to the observed dramatic enhancement of the isotope effect\cite{isotope}, as coupling to the slower lattice degree of freedoms is more effectively for heavier pairs.
Similarly, together with mass enhancement, the proximity to the incoherent local $p$-wave [c.f.:Fig.~\ref{fig:fig3}(c)] allows the observed increase of residual specific heat\cite{Cv}.
Given their finite amplitude along the $d$-wave nodal directions [c.f.:Fig.~\ref{fig:fig3}(d)(f)], the enhanced fluctuation to local $p$-wave pairs also can explain the recently observed pseudogap along the nodal direction~\cite{nodal_pgap} in heavily underdoped samples.
At $\delta < \delta_c$, the infinite degeneracy of the incoherent $p$-wave along the antinodal directions [c.f.:Fig.~\ref{fig:fig3}(a)], with their infinite mass and unusually enhanced scattering, gives a new paradigm to the insulating~\cite{MIT} glassy~\cite{glass} electronic structure and the non-fermi-liquid transport~\cite{T_linear} 
Our result suggests that the system is glassy not only in the spin channel, but also in the charge and pairing channel as well.
Obviously, our theory is consistent with the observed charge 2$e$ quanta across the superconducting-insulating transition~\cite{charge2e}, which raised the serious issue "How can a system of charge 2$e$ bosons be insulating?  If it is just Anderson localization, how can $\delta_c$ not present strong sensitivity to disorder?"
Our result provides a long-sought disorder-insensitive alternative paradigm.
Finally, it is curious to notice, across $\delta_c$, the same 45$^{\circ}$ rotation in the directions of the dominant hopping, the nodal structure of local pairs, and the observed stripe correlation~\cite{stripe_rotation}.

In conclusion, we demonstrate that all the key features of superconductivity in the underdoped cuprates can be described quantitatively in the strong binding limit, without use of any free parameter.
The $d$-wave symmetry is found to originate from the renormalized kinetic energy, and the observed superconductivity can be understood as a superfluid of a dilute real-space hole pairs.
Our result explains the lack of superconductivity at $\delta < \delta_c$ due to quantum fluctuation associated with incoherent local $p$-wave pairs.
In the underdoped regime, a large effective mass enhancement of the hole pairs is found responsible for the observed weak phase stiffness.
Finally, the observed $\delta=5.2\%$ QCP is found dictated by the divergence of the effective mass of the hole pairs, which also make sense the dramatic reduction of diamagnetic response (the Nernst effect) near the QCP.
These successes support strongly a simple description of bosonic condensate for the underdoped cuprates and enable further reconciliation of seemingly contradicting experimental conclusions in the field.

We acknowledge useful discussions with Maxim Khodas and Chris Homes, and comments from Alexei Tsvelik and Weiguo Yin.  This work was supported by the U.S. Department of Energy, Office of Basic Energy Science, under Contract No. DE-AC02-98CH10886.


\begin{thebibliography}{10}
\bibitem{pseudogap} The higher-energy pseudogap renormalizes the kinetics signifiantly, and in turn affects the superfluid phase stiffness.  This is to be distinguished from the reduced superfluid density due to pseudogap surpressed Fermi surface spectral weight in the weak-coupling scenario.
\bibitem{Tc_determination}
$T_c$ is determined by thermally exhausting all the pairs $\int_\mu d\omega DOS(\omega) / (e^{\beta(T_c) \omega}-1)=\delta$.
\end{thebibliography}
\end{document}